\documentclass[a4paper,11pt]{article}
\usepackage[utf8]{inputenc}
\usepackage{jcappub} % for details on the use of the package, please see the JINST-author-manual
\usepackage{graphicx}
\usepackage{hyperref}
\usepackage{xspace}
\usepackage{aas_macros}
\newcommand{\lat}{{\it Fermi}-LAT\xspace}

\newif\ifrev
\revtrue              % <-- Set to \revfalse to turn OFF highlighting

\newcommand{\rev}[1]{%
  \ifrev
    #1%% \textbf{#1}%
  \else
    #1%
  \fi
}

\arxivnumber{2604.19916} % Only if you have one
\title{\boldmath Search for Anisotropic Pair Halos Associated with Blazar Jets}

\author[a]{Ao Zhang,\ }
\author[b]{Wenlei Chen,\ }
\author[a]{and Manel Errando}
\affiliation[a]{Department of Physics, Washington University in St. Louis, 1 Brookings Drive, St. Louis, MO 63130, USA}
\affiliation[b]{Department of Physics, Oklahoma State University, 145 Physical Sciences Building, Stillwater, OK 74078, USA}

\emailAdd{ao.z@wustl.edu}

\abstract{
The origin of intergalactic magnetic fields (IGMFs) remains one of the key open questions in cosmology. Gamma-ray pair halos produced by electromagnetic cascades from TeV-emitting blazars provide a powerful indirect probe of these fields. In this work, we present a novel search for pair halos that explicitly exploits their expected anisotropic morphology, aligning with the projected orientation of blazar jets on the sky. Using a Monte Carlo framework to model the spatial distribution of cascade emission, we identify an optimal sample of 21 high-synchrotron-peaked BL Lac objects with well-constrained jet position angles from radio interferometry. By rotating and stacking \textit{Fermi}-LAT observations of these sources along their jet directions, we enhance sensitivity to anisotropic extended emission that would be diluted in traditional orientation-agnostic analyses. 

Applying a likelihood analysis to the combined dataset, we find evidence for a non-zero IGMF, excluding the null hypothesis at $3.8\sigma$ level and obtaining a best-fit field strength of $B_0 = 2.8 \times 10^{-16}\,\mathrm{G}$, with a $99\%$ confidence interval of $0.9 \times 10^{-16}\,\mathrm{G} < B_0 < 8.9 \times 10^{-16}\,\mathrm{G}$. Our result is consistent with previous constraints from spectral, spatial, and temporal studies, while demonstrating that incorporating anisotropic information provides a significant gain in sensitivity. This approach opens a new avenue for probing intergalactic magnetism and highlights the potential of future high-angular-resolution gamma-ray observations to directly image pair halos and map magnetic fields in cosmic voids.}

\begin{document}
\maketitle
\flushbottom
\section{Pair haloes from TeV-emitting blazars}

Magnetic fields in galaxies and clusters, with intensities that typically range between 1 and 10~$\mu$G, are observed using the Faraday rotation and the Zeeman splitting effect \cite{1999ARAA..37...37K,Carilli_2002}. These fields are believed to evolve from weaker ``seed'' fields through the ``$\alpha$--$\omega$'' dynamo mechanism and turbulent plasma motions during galaxy formation \cite{1950ZNatA...5...65B, Kulsrud_2008}. However, the origin of these weaker ``seed'' fields remains uncertain, as no widely accepted theory explains their formation \cite{Grasso_2001, 2002RvMP...74..775W, Kulsrud_2008}. Measuring these weak fields could provide valuable insights into their origins.

\begin{figure}[tbp]
\centering
\includegraphics[width=0.8\textwidth]{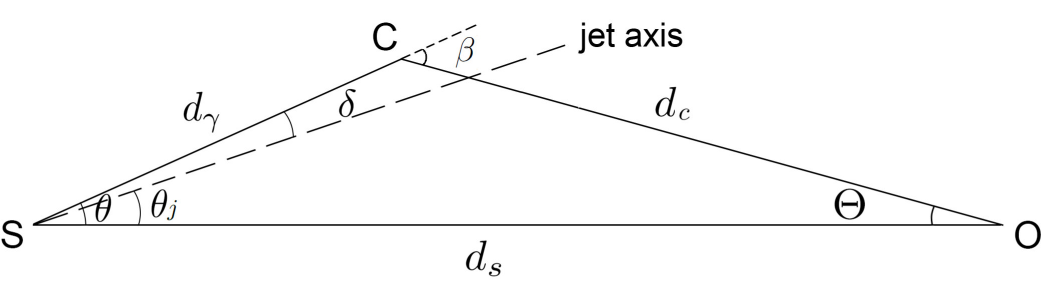}
\caption{Geometry of the pair cascade interaction. The gamma-ray source it at S and observer at O. $\Theta$ is the angular distance from the source location at which the secondary GeV photons will be observed, and $\beta$ is the integrated deflection angle of the electron or positron. Adapted from \cite{chen_2018}.}
\label{geometry}
\end{figure}

The intergalactic medium may retain the original structure of the weak seek field and therefore can be used to investigate its strength. However, the absence of detectable intergalactic magnetic fields (IGMFs) through Faraday rotation from radio sources or in the cosmic microwave background indicates that these fields are extremely weak \cite{Vachaspati_2021}. Constraints based on these methods have placed an upper limit at $B\lesssim 1\,\text{nG}$ \cite{Planck_2016, Pshirkov_2016}. Indirect detection methods involve using gamma-ray observatories like the Fermi Large Area Telescope (\lat) or the Cherenkov Telescope Array Observatory, which could observe evidence for electron-positron pair halos forming around primary gamma ray sources \cite{Neronov_2009}. 
%Unlike the GeV photons which are mostly transparent ($\tau\lesssim0.1$) through the intergalactic media at low redshift ($z \lesssim 0.2$), TeV photons emitted by active galactic nuclei (AGNs) interact with the extragalactic background light (EBL), leading to pair production (\cite{Domnguez_2010}).
Unlike GeV photons, which remain largely unattenuated ($\tau\lesssim0.1$) when propagating through the intergalactic medium, %at low redshift ($z \lesssim 0.2$), 
photons with primary energy \rev{ $10\,\text{TeV} \gtrsim E_{\gamma 0} \gtrsim 1.0\, \text{TeV}$} from active galactic nuclei (AGNs) interact with the extragalactic background light (EBL), leading to pair production \cite{1994ApJ...423L...5A,Domnguez_2010}. 
The mean free path for pair production of a gamma-ray photon with primary energy $E_{\gamma 0}$ propagating from a source at redshift $z$ is given by:
\begin{equation}
D_{\gamma}\approx 80\,\text{Mpc}\,\frac{\kappa}{(1+z)^2}\left(\frac{E_{\gamma 0}}{\text{10\,TeV}}\right)^{-1}
\end{equation}
where $\kappa \sim 1$, \cite{Neronov_2009}.  %, and is typically on the order of $10^{2}$ Mpc. 
The resulting electron-positron pairs are deflected in the presence of the IGMF and subsequently scatter photons from the cosmic microwave background to energies of a few GeV at the redshift of $z_{\gamma\gamma}$ with a mean free path for Compton scattering $D_{IC}\approx 1.2\,\text{kpc}\, (1+z_{\gamma\gamma})^{-3}$\cite{Neronov_2009}. %During the inverse Compton scattering process, the electrons and positrons lose energy, 
As a result of Compton scattering, electrons and positrons lose energy, cooling to half their initial energy over a cooling distance $D_e$. Assuming that the Larmor radius of the electrons is $R_{L}\gg D_{e}$, the deflection angle $\beta$ can be approximated as $\beta = {D_e}/{R_L}$, where $R_L$ %is the Larmor radius and 
is inversely proportional to the magnetic field strength. The geometry of this process is shown in Figure~\ref{geometry}. 

%TODO: include pair halo geometry here. 

The morphology of pair halos is influenced by the strength and coherence length of the intergalactic magnetic field. When the IGMF is relatively strong, $B\gtrsim 10^{-12}$\,G, secondary gamma rays will be distributed isotropically because the Larmor radius of the charged particles is comparable or smaller than the cooling length for pairs. In contrast, for weaker magnetic fields, the Larmor radius is much larger than the cooling length, leading to an extended, offset GeV emission from the source that gamma-ray observatories could resolve. Another critical factor is the coherence length $\lambda$ of the IGMF. For %coherence length comparable or smaller than the cooling distance
$\lambda \lesssim D_e \sim 10$ kpc, charged particles become diffusive and the simple geometry in Figure~\ref{geometry}
does not apply . Conversely, if the magnetic field is uniform in scales $\lambda \gtrsim D_\gamma\sim 100$\,Mpc, the observed secondary gamma rays will be anisotropic and perpendicular to the direction of the magnetic field \cite{Long_2015,2016ApJ...832..109B,Batista}. %To possibly detect pair halo signals around the sources, the coherence length is in the range of $30 \text{ kpc}\lesssim \lambda \lesssim 30 \text{ Mpc}$, 
The range of possible values of the magnetic field coherence length in the intergalactic medium is largely unconstrained, with the theoretical range spanning $10^{-11} \text{Mpc }\lesssim \lambda \lesssim R_{H}$, where $R_{H}$ is the Hubble radius \cite{Grasso_2001, Neronov_2009}.

Recent studies have placed constraints on the strength of IGMF by investigating the spectral, spatial extension (pair-halo) and the time-delayed (pair-echo) emission of secondary gamma rays from the TeV emitters. Based on the non-observation of secondary GeV signals in the spectrum of TeV sources, a lower limit, $B\geq 3\times 10^{-16}$ G, can be set to the IGMF \cite{Neronov_2010}. Additional studies have investigated extended gamma-ray emission from blazars using different observatories. From the \lat dataset,  a hint of a pair halo signal was reported by stacking the emission from 24 BL~Lac-type blazars at low redshift, constraining the strength of IGMF $10^{-15}\,\text{G} \leq B \leq10^{-17}$\,G \cite{Chen_2015}. Other constraints have been derived from observations at higher energies with imaging atmospheric Cherenkov telescopes such as MAGIC, H.E.S.S., and VERITAS \cite{Aleksic_2010, Hess_2014, Archambault_2017}. Searches for time-delayed secondary emission from blazars and gamma ray bursts have also been conducted, %. This method has also provided 
providing constraints on IGMF ranging $10^{-16}$ G to $10^{-19.5}$ G \cite{wang_2020, Magic_2023, mirabal_2023, huang_2023, vovk_2024}. See \cite{2021Univ....7..223A} for a comprehensive review of current limits and a comparison of different techniques.

\section{Monte Carlo model for pair haloes} \label{model}
We developed a Monte Carlo pair halo model that calculates the spatial distribution of secondary gamma rays by mixed ray-tracing \cite{chen_2018}. This extended pair halo component depends on the jet profile and viewing angle, the redshift of the source, the spectrum of the interacting low-energy photon fields, and the properties of the IGMF. To reduce the computational cost, we assumed the same jet geometry for all of our sources and simulate all the primary gamma rays at a fixed energy. We %used $1^{\circ}$ as the jet opening angle and 
modeled the jet profile as a two-dimensional Gaussian distribution with $1^{\circ}$ full width at half maximum (FWHM). \rev{We adopt this jet opening angle based on a study of 135 MOJAVE AGNs, which measured a median intrinsic jet opening angle of $1.3^{\circ}$ \cite{Pushkarev2017}.} Because we are selecting sources whose jet axes are not fully aligned to our line of sight, the jet viewing angle is fixed at $\theta_j=0.5^{\circ}$, meaning the observer is at the edge of the jet cone. The IGMF in our model is random and non-helical with coherence length of 1\,Mpc. This choice of coherence length is reasonable for our pair halo search because it is smaller the mean free path of pair production for TeV photons ($\sim 100$ Mpc) and larger than the average cooling distance of energetic pairs ($\sim 30$ kpc) \cite{Neronov_2009}. The simulation code we set up \rev{was based on the analytic framework from \cite{Neronov_2009}, and simulations from \cite{Neronov_2010_2}. We adopted EBL model developed in \cite{Franceschini2008} for calculating the pair production processes.} Examples of the output of our simulations are shown in Figure~\ref{ph_sim}.

%Include simulated pair halo example here? OR in Appendix... 

\section{Selection of an optimal blazar sample for IGMF searches}
%\subsection{Source Selection}
%select sources based on redshift and P.A.
% In order for the primary photons to pair produce, sources need to emit TeV gamma rays and high-synchrotron-peaked BL Lacs (HBLs) are the only known class of blazars to emit TeV gamma rays (cit). 

To maximize our sensitivity to probe the IGMF via the detection of a pair halo signal, we run our search on a sample of selected high-synchrotron-peaked BL Lacs (HBLs). HBLs are characterized by a synchrotron peak frequency above $10^{15}\,\text{Hz}$ %a known class of TeV emitters, 
and are the dominant population of TeV sources in the extragalactic sky. %with synchrotron peak frequency above $10^{15} \text{Hz}$. 
In addition, the ability to detect pair halo signals depends on the distance or the redshift of the source. For objects at $z \lesssim 0.02$, the distance to the source becomes comparable or smaller than the mean free path for pair production of TeV photons on the EBL ($\sim 100\,\text{Mpc}$, \cite{Neronov_2009}), resulting on a reduced yield of detectable secondary photons that would constitute the pair halo signal. % If the source is too close, little TeV photons can undergo pair productions with the EBL as the mean free path is around 100\,Mpc \cite{Neronov_2009}, resulting a smaller chance of detecting secondary gamma rays. 
For distant sources, the halo would fully develop close to the source but at a large distance from the observer, resulting on a very compact spatial extension of the pair halo that would %If the source is too far, the deviated angle of the secondary gamma rays will be so small that they 
fall inside the point spread function (PSF) of GeV observatories. %containment circle of the instrument psf and therefore, are undistinguishable from the primary gamma rays. 

\begin{figure}[tbh]
\centering
\includegraphics[width=0.75\textwidth]{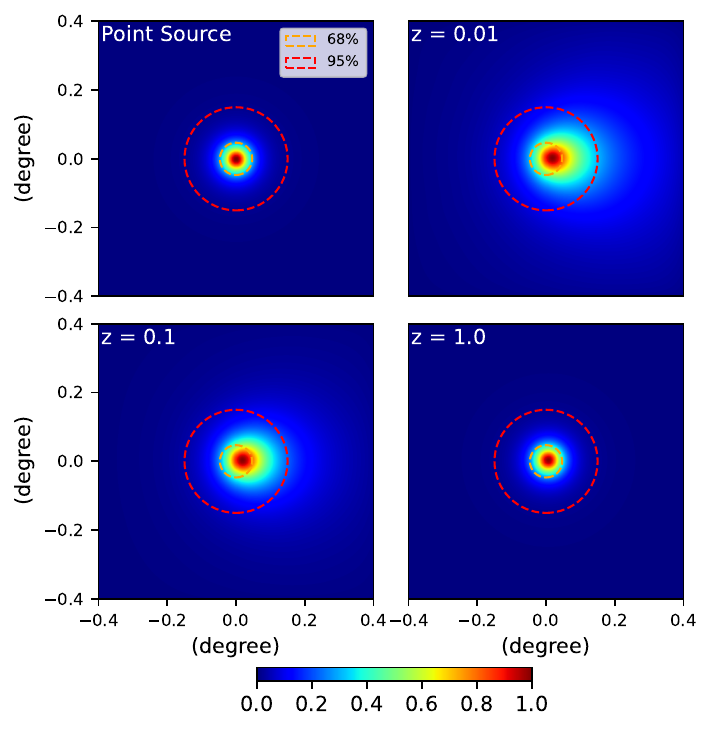}
\caption{The top left panel shows the 2-D instrumental point spread function of \lat. The orange and red dashed circles represent the $68\%$ and $95\%$ containment radii, respectively. The remaining panels show the simulated spatial distribution of secondary $30$\,GeV photons originating from pair cascades from a source at $z = 0.01$, $0.1$, $1.0$ and assuming an IGMF with intensity %overlapped with the psf containment circles of Fermi-LAT at 
$B_0 = 10^{-15} G$. The jet profile is modeled as a 2-D Gaussian distribution with jet opening angle of $1^{\circ}$ and the viewing angle is $0.5^{\circ}$.}
\label{ph_sim}
\end{figure}

\begin{figure}[tb]
\centering
\includegraphics[width=0.75\textwidth]{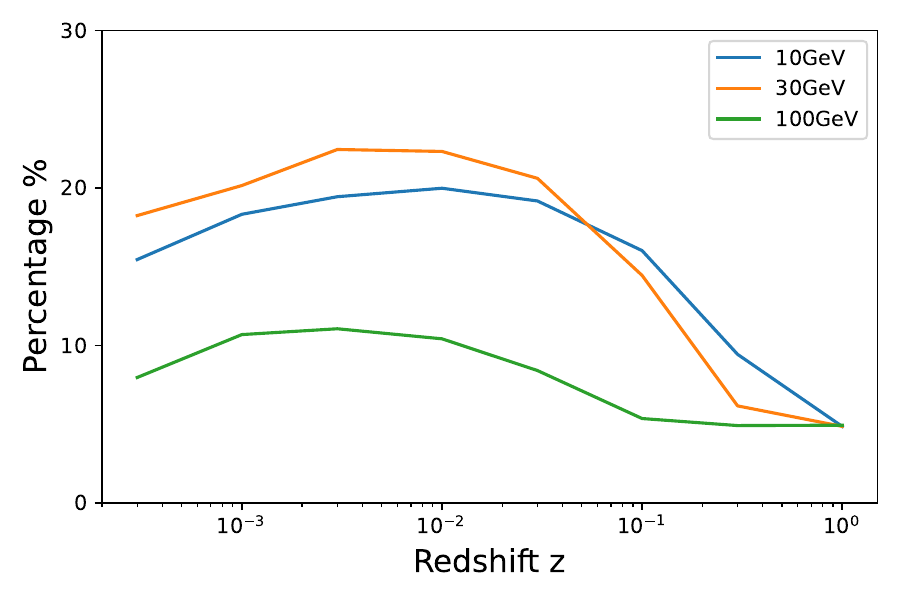}
\caption{Percentage of $10$\,GeV, $30$\,GeV and $100$\,GeV secondary gamma-ray photons falling outside the $95\%$ containment circle of the \lat PSF as a function of redshift of the simulated primary gamma-ray source. Examples of the spatial photon distributions at 30\,GeV are shown in Figure~\ref{ph_sim}. }%instrumental psf as a function of redshift.}
\label{percentage}
%\vspace{0.5cm}
\end{figure}

To find an optimal redshift range that maximizes the expected pair halo signal, we use our pair halo model %\cite{chen_2018} 
to simulate the $2D$ spatial distribution of secondary gamma rays, and calculate the percentage that fall outside the $95\%$ containment angle of the \lat PSF. Assuming a module of the  IGMF of $10^{-15}$\,G, the spatial distribution of $30$\,GeV secondary gamma rays at redshift of $z=0.01$, $0.1$, $1.0$ is shown in Figure~\ref{ph_sim}. The choice of benchmark magnetic field intensity is somewhat arbitrary but in line with  recent constraints \cite{Aharonian_2023}. % We used $10^{-15}$ G for our simulation because recent paper has suggested the IGMF to be $B>7.1 \times 10^{-16}$ G \cite{Aharonian_2023}. In addition, we have assumed that the observer is at the edge of a jet cone, in which the pair halo signal will have an anisotropic signature. 
%
%%TODO: Add more information here, how many particles simulated. Include simulated reshift graph
%
Based on this simulation exercise, %shown in Figure~\ref{ph_sim}, 
we calculated the percentage of secondary gamma rays that fall outside the $95\%$ containment circle of instrumental PSF. % as a function of redshift for three different energies of secondary $\gamma$ rays. 
%This calculation does not account for the brightness of secondary pair halos. However, considering brightness, we set the lower redshift limit to 0.03 because TeV photons emitted by HBLs at lower redshifts rarely undergo pair production, given that the mean free path for pair production is $\sim100$ Mpc \cite{Neronov_2009}. 
As seen in Figure~\ref{percentage}, the percentage of secondary photons that would be detectable as a resolved, spatially extended gamma-ray emission component declines steeply at $z \gtrsim 0.1$. Therefore, we restrict our sample of TeV emitting HBLs to the range $0.03 \leq z \leq 0.15$. \rev{Restricting our sample to $0.03 \le z \le 0.15$ optimizes sensitivity by selecting targets where the cascade yield is high and a significant fraction of secondary photons extend beyond the Fermi-LAT PSF.}
%Therefore, we think that the redshift of HBLs between 0.03 and 0.15 is ideal to search for pair halo signal. 
We used the fourth \lat AGN catalog \cite[4LAC, ][]{2022ApJS..263...24A} to compile a list of LAT-detected AGN within this redshift range. To select HBL objects that are likely to produce TeV emission, we chose sources with estimated synchrotron peak frequency $\nu_{\text{syn}}>10^{15}$\,Hz. In addition, to minimize any potential contamination to the pair halo signal from unrelated nearby gamma-ray sources, we excluded from our analysis any sources that had other GeV sources listed in the third \lat catalog of high-energy sources \cite[3FHL, ][]{2017ApJS..232...18A} within a $1^{\circ}$ radius. %that are significantly detected in the third \lat catalog of high-energy sources (3FHL). 
%I included an explanation of radio observation here. 

When the jet from a blazar is slightly misaligned with respect to our %sources have jets slightly off-aligned with our 
line-of-sight, the pair halo will develop along the projected direction of the jet in the sky (Figure~\ref{geometry}, \cite{Neronov_2010_2}). %as shown in Figure~\ref{geometry}, the secondary $\gamma$ rays tend to develop along the jet position angles with the presence of IGMF. 
To take advantage of this asymmetry and use it to increase the signal-to-noise ratio of our search, %herefore, 
we only select HBLs whose jet position angles are known from radio interferometry observations. That way, we can % so that we can 
rotate the counts map to align the jet position angles from multiple sources and stack their pair halo signals along the jet direction. With data from the Very Long Baseline Array (VLBA), we are able to constrain the jet position angle using radio observation. 
% Finally, to minimize any potential contamination to our pair halo signal, we excluded any HBLs that have other GeV sources within $1^{\circ}$ that are significantly detected in the third \lat catalog of high-energy sources (3FHL). 

%TODO: Recalculate f_halo using new spectrum

\begin{table}[bht]
\centering
\resizebox{\textwidth}{!}{
\begin{tabular}{c c c c c c c c c c}
\hline
Blazar name & Redshift & P.A. ($^\circ$) & $\alpha$ & $f^{LE}_{5\text{TeV}}$ & $f^{LE}_{10\text{TeV}}$& $f^{HE}_{5TeV}$  &$f^{HE}_{10TeV}$& $\text{TS}_{5\, TeV}$ &$\text{TS}_{10\, TeV}$\\
\hline
RBS 0030 &$0.0948$& $191.5^{*}$  & $1.9046$
& $0.11$ &$0.17$& $0.05$ &$0.09$&$0.04$&$0.07$\\
PMN J0152+0146 &$0.08$& $229.1^{*}$  & $1.9987$
& $0.09$ &$0.15$& $0.06$ &$0.1$& $0.1$&$0.13$\\
TXS 0210+515 &$0.049$& $70.4$ & $1.8674$
& $0.11$ &$0.18$& $0.05$ &$0.12$& $2.23$&$3.74$\\
1ES 0229+200 & $0.1396$& $161.4^{*}$  & $1.7518$
& $0.21$ &$0.32$& $0.1$ &$0.18$& $0.13$&$0.25$\\
TXS 0518+211 & $0.108$& $264.5$ & $1.9518$
& $0.07$ &$0.19$& $0.03$ &$0.13$& $0.13$&$0$\\
PKS 0548-322 & $0.069$& $7.6^{*}$  & $1.8294$
& $0.15$ &$0.23$& $0.07$
 &$0.15$& $0.7$&$1.23$\\
1H 0658+595 & $0.125$& $203.2^{*}$  & $1.7033$
& $0.27$ &$0.36$& $0.13$ &$0.21$& $0.48$&$0.64$\\
1ES 0806+524 & $0.138$& $68.2$ & $1.8825$
& $0.14$ &$0.2$& $0.08$ &$0.12$& $0.49$&$0.76$\\
1RXS J101015.9-311909 & $0.1426$& $15.4^{*}$  & $1.7653$
& $0.21$ &$0.32$& $0.09$ &$0.2$& $0.17$&$0.26$\\
Mrk 421 & $0.031$& $330.3$ & $1.7846$
& $0.14$ &$0.2$& $0.07$ &$0.13$& $5.25$&$6.62$\\
Mrk 180  & $0.045$& $106.9$ & $1.8277$
& $0.13$ &$0.19$& $0.06$ &$0.12$& $1.28$&$1.7$\\
H 1426+428 & $0.129$& $350.5^{*}$  & $1.6587$
& $0.29$ &$0.44$& $0.13$ &$0.27$& $0.23$&$0.42$\\
TXS 1515-273 & $0.1284$& $89.2$ & $2.0362$
& $0.05$ &$0.14$& $0.04$ &$0.09$& $0.59$&$0.84$\\
1ES 1727+502 & $0.055$& $286.8$ & $1.7846$
& $0.14$ &$0.26$& $0.06$ &$0.19$& $0.89$&$1.3$\\
1ES 1741+196 & $0.084$& $70.1$ & $1.9216$
& $0.09$ &$0.16$& $0.05$ &$0.1$& $1.7$&$2.67$\\
1H 1914-194 & $0.137$& $332.8$ & $1.9452$
& $0.09$ &$0.17$& $0.04$ &$0.12$& $0.06$&$0.08$\\
1ES 1959+650 & $0.048$& $124.1^{*}$  & $1.815$
& $0.12$ &$0.22$& $0.06$ &$0.16$& $2.05$&$2.9$\\
PKS 2005-489 & $0.071$ & $216.6^{*}$  & $1.8408$
& $0.14$ &$0.2$& $0.06$ &$0.11$& $0.39$&$0.6$\\
PKS 2155-304 & $0.116$ & $168.5^{*}$  & $1.8573$
& $0.14$ &$0.23$& $0.07$ &$0.15$& $3.98$&$5.34$\\
B3 2247+381 & $0.1187$& $291.7^{*}$  & $1.7696$
& $0.2$ &$0.34$& $0.1$ &$0.21$& $2.57$&$3.46$\\
1ES 2344+514 & $0.044$& $140.1$ & $1.8325$
& $0.12$ &$0.19$& $0.07$ &$0.1$& $4.51$&$6.58$\\
\hline
\end{tabular}
}
\caption{List of targets selected for our analysis, ordered by R.A. The jet position angles without asterisks are calculated by a weighted average of different components of the jet using the MOJAVE database \cite{Lister_2019}. P.A. with asterisks are calculated using image moment analysis. The spectrum for each source is assumed to be $dN/dE\propto E^{-\alpha}\ \exp{-{E}/{E_c}}$ %with $E_c = 5\,\text{TeV}$ and $E_c = 10\,\text{TeV}$
with reported spectral index $\alpha$ from the Fermi Point Source Catalog, 4FGL-DR4 \cite{2022ApJS..260...53A}.}
\label{bl_list1}
\end{table}

%TODO: put self measured blazars from VLBA websites. Double check the position angles. Maybe a more scientific way to measure the position angle? 

%Included a list of HBLs in which the pa are calcualted by moment analysis.
%TODO: double check the blazar names. 

%TODO: add the search process..

% \begin{table}[htbp]
% \centering
% \begin{tabular}{c c c}
% \hline
% Blazar name & Redshift & P.A. ($^\circ$)\\
% \hline
% PKS 0548-322 & 0.069 & 7.6\\
% RBS 0030 & 0.0948 & 191.5\\
% PMN J0152+0146 & 0.08 & 229.1\\
% 1ES 1959+650 & 0.048 & 124.1\\
% 1ES 0229+200 & 0.1396 & 161.4\\
% 1H 0658+595 & 0.125 & 203.2\\
% 1RXS J101015.9-311909 & 0.1426 & 15.4\\
% H 1426+428 & 0.129 & 350.5\\
% PKS 2155-304 & 0.116 & 168.5\\
% B3 2247+381 & 0.1187 & 291.7\\
% PKS 2005-489 & 0.071 & 216.6\\
% \hline
% \end{tabular}
% \caption{A list of selected targets from VLBA observations. The jet position angles are calculated using moment analysis of the image. \label{bl_list2}}
% \end{table}

%TODO: detailed explanation about image moment analysis, include an image here? OR in appendix?
By applying the above mentioned conditions, we have find 21 HBLs that meet all our selection criteria. Among them, ten sources have jet position angles available from the MOJAVE program \cite{Lister_2019}. %The jet position angles of 
For the remaining objects, we used archival VLBA data and calculated their jet position angle %of VLBA observations using 
via image moment analysis \cite{piner_2018,TANAMI}. \rev{While 18 of the 21 targets lie at high Galactic latitudes $|b|>10^{\circ}$, 3 targets reside at low latitudes down to $|b|\approx8.7^{\circ}$. However, restricting the energy range to $E > 30\text{ GeV}$ and applying the \textit{P8R3\_ULTRACLEANVETO} event class effectively suppresses Galactic diffuse emission, ensuring a background-free analysis across all selected fields.} The 21 selected sources are listed in Table~\ref{bl_list1}.

\section{Processing and analysis of \lat data} \label{data}
\begin{figure}[tb]
\centering
\includegraphics[width=.48\linewidth]{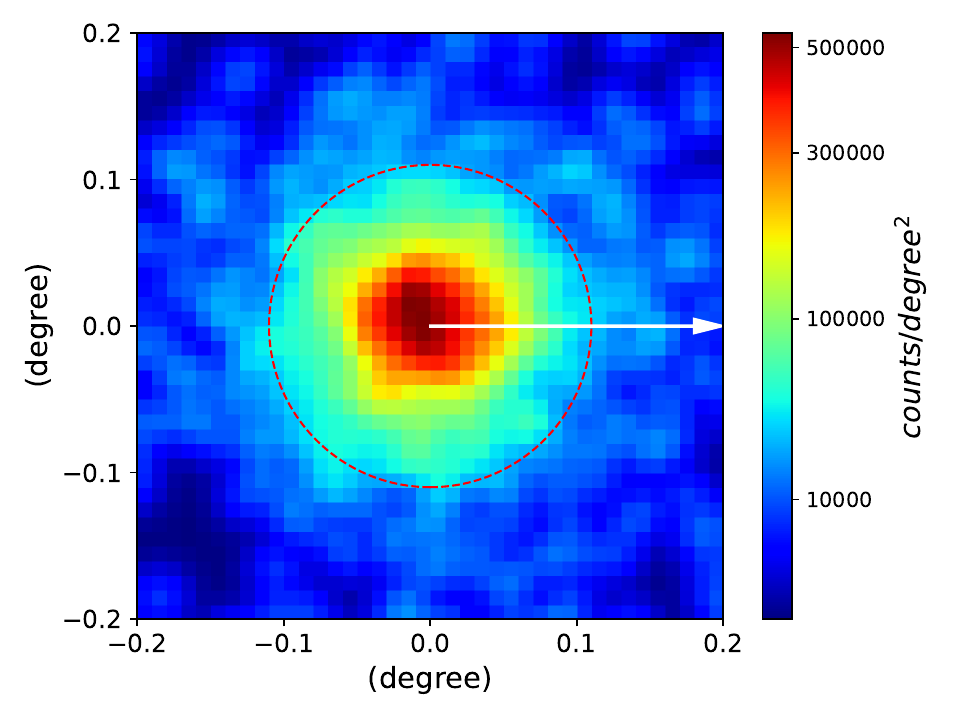}
\includegraphics[width=.48\linewidth]{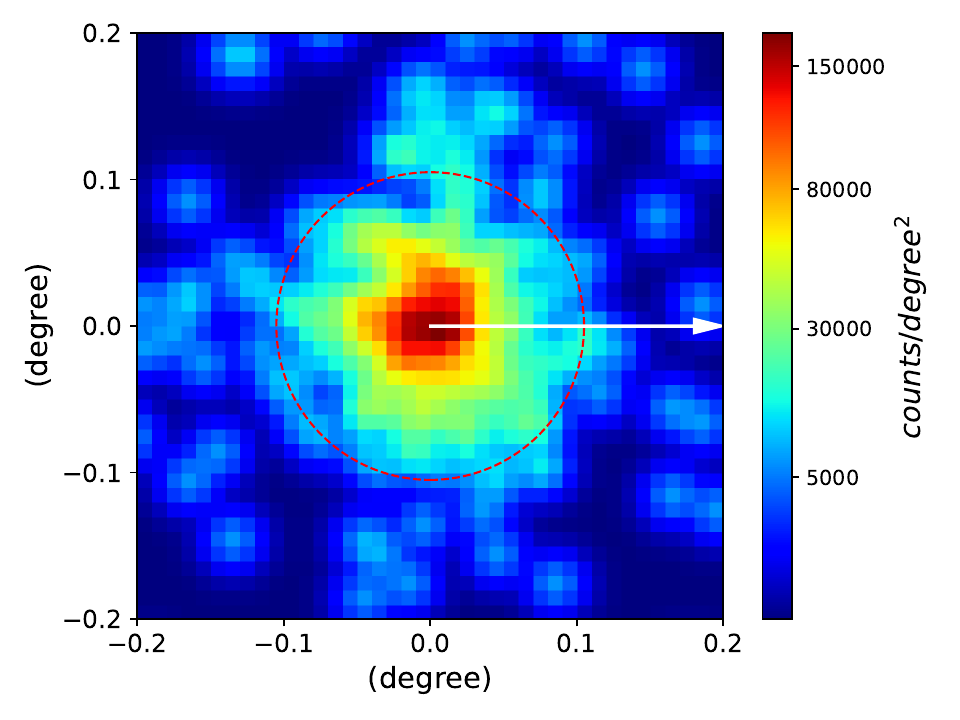}
\caption{Stacked sky maps of the 21 sources in our sample, rotated to have their projected radio jet aligned with the positive direction of the X axis (white arrow) in two energy bins:. $30-100$\,GeV (left) and $100-300$\,GeV (right). The red circle shows the $68\%$ containment radius of \lat. The map is smoothed by a Gaussian filter with $\sigma = 0.005 ^\circ$ using \textit{scipy.ndimage.gaussian\_filter}. This figure is for illustration only and is not used for joint likelihood ratio test.}
%\vspace{-0.5cm}
\label{stacked_map}
\end{figure}

%TODO: CHECK the VLBA jet core position! 
To search for a potential pair halo signal, we selected \lat Pass 8 data from mission week 9 to week 835. To minimize background emission and noise, we used the cleanest event class \textit{P8R3\_ULTRACLEANVETO} and selected photons with energy above 30 \,GeV. % energy filter when selecting photons from event files. 
We applied recommended quality cuts to our photon sample: zenith angle $<90^{\circ}$, DATA\_QUAL$>0$, and LAT\_CONFIG == 1. Events from both conversion layers (front and back) are included in our analysis. Because our pair halo model only simulates gamma rays at a fixed energy, we divide the gamma ray events into two energy bins: $30-100$\,GeV, and $100-300$\,GeV. The sky map of each source is prepared as a $80\times80$ pixel map with $0.01^\circ \times 0.01^\circ$ pixels, centered at the VLBA position of the source. %This choice of the map is reasonable because 
The $68\%$ containment angle of the \lat PSF is around $0.1^\circ$ at 30\,GeV. We then rotate the sky map around the jet core position to align the projected direction of the jet in the sky with the positive direction of the %position angle with the positive 
x-axis using the \texttt{axisrot} function in the analysis tool \texttt{gtbin} %, with the most updated software version: 
v11r5p3. The milliarcsecond localization precision of the radio core from radio interferometry observations %In addition, the radio measurement of the positions of the sources has milliarcseconds precisions which 
prevents any anisotropic signals due to offsets between our assumed location of the jet core and its true location. %misalignment of the sources. We 
The sky map that results from stacking the rotated counts maps of 21 sources in our sample  is shown in Figure~\ref{stacked_map}.

\section{Detection of a non-zero intergalactic magnetic field}
To perform a quantitative test for the presence of a potential pair halo signal, we conducted a statistical analysis in which we assumed that the observed gamma rays in the \lat data originate from three different emission components: direct emission from the jet that would manifest as a point source, emission from the secondary extended pair halo, and diffuse background emission. We then performed a likelihood ratio test between the hypothesis in which the observed count maps contain a pair halo signal generated by an IGMF with field strength $B_0$, and the null hypothesis in which $B_0 = 0\,\text{G}$ and no extended pair halo signal would be produced. 
Given a dataset $D$, the joint likelihood function can be given by: 
\begin{equation}
\label{TS_eq}
L(x|D) = P\left(\sum N_{\text{off},ijk}|\sum \mu_{ijk}(x))\right)\times \prod_i\prod_j\prod_k P\left(N_{\text{on},ijk}|\lambda_{ijk}(x)\right)
\end{equation} 
%where $x = [B_0, {f^{\text{halo}}_{ij}}, \theta_i, \{N_{\text{off},ijk}\},\{N_{\text{on},ijk}\}]$ 
where $x = [B_0, {f^{\text{halo}}_{ij}}, \{N_{\text{off},ijk}\},\{N_{\text{on},ijk}\}]$ 
defines the parameter space for the $i^{th}$ source at $j^{th}$ energy bin at $k^{th}$ spatial pixel. %$B_0$ is the strength of the IGMF. 
$f^{\text{halo}}_{ij}$ is the fractional intensity of the pair halo emission with respect to the total number of observed photons, which includes the direct primary emission from the source. 
%$\theta_i$ is the jet viewing angle, in which we fixed at $0.5^\circ$. 
$N_{\text{on},ijk}$ and $N_{\text{off},ijk}$ are the observed source and background counts, while $\mu_{ijk}$ and $\lambda_{ijk}$ are the expected source and background counts from the simulated pair halo. As discussed in Section~\ref{data}, the background emission is minimized by applying an energy filter to select gamma rays above 30\,GeV. It is safe to assume that our observational data is free of background events and we therefore set $ N_{\text{off},ijk} = 0$. %Therefore, we only need to consider the point source and pair halo contributions to the total observed emission. 

%Because the likelihood ratio test is a ratio of a non-zero pair halo contribution against a zero pair halo contribution, we need to quantify the fractional component of the secondary cascade emissions with respect to the primary gamma ray emissions from the source. 

A key requirement of our statistical test is to quantify the intensity of the secondary pair halo emission, characterized in our model by the parameter $f^\text{halo}$.
To provide a quantitative estimate for $f^\text{halo}$ for each source and in each of the two energy ranges we explore,  
%Because this parameter is essential to our likelihood analysis, we decided to use a more sophisticated and well-developed Monte-Carlo framework, 
we used the \texttt{CRPropa}  Monte Carlo simulation framework \cite{Alves_Batista_2022}. %Unlike our code which simulates particles at one energy value, 
\texttt{CRPropa} takes the intrinsic, primary gamma-ray spectrum of a source as input and simulates the interactions of gamma-ray photons (pair production and Compton scattering) as they propagate through the intergalactic medium, which is characterized by its radiation content and magnetic field properties. % under multiple photon fields, and traces secondary particles with user-defined magnetic fields. 
Furthermore, \texttt{CRPropa} allows secondary gamma rays to undergo pair production and Compton scattering with ambient photon fields multiple times, unlike our simplified model described in Section~\ref{model} that only allows a single pair production and Compton scattering sequence.  %which would be a much more realistic scenario compared to our one-time pair production process for each gamma ray %\cite{Alves_Batista_2022}. 

\begin{figure}[tbh]
\centering
\includegraphics[width=0.7\textwidth]{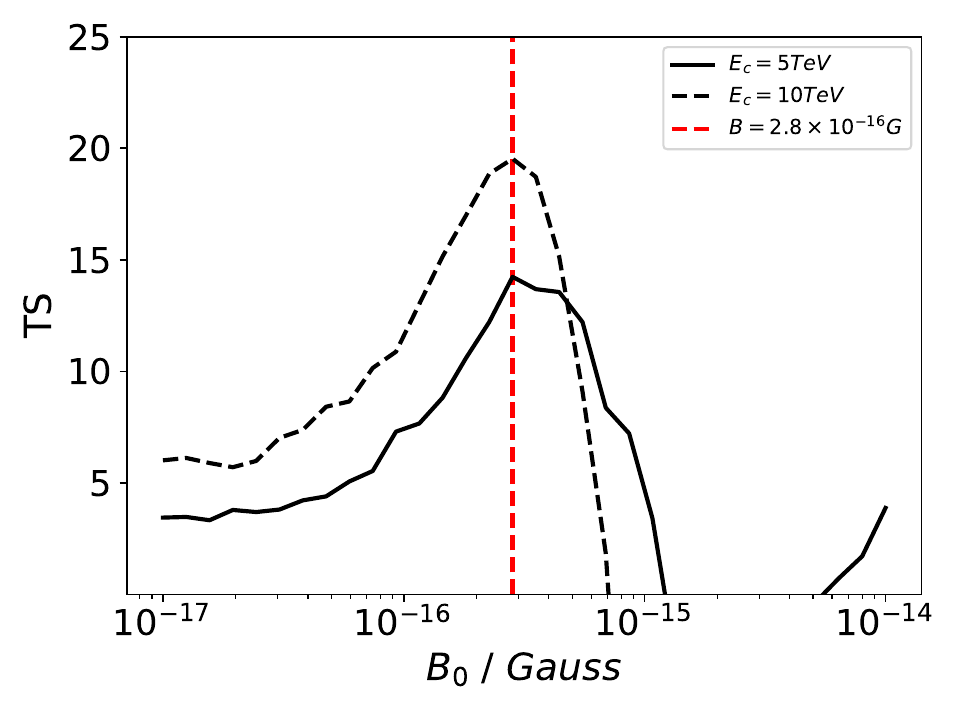}
\caption{Combined test statistic (TS) of the model containing pair halo emission in the \lat data from our 21 selected sources as a function of the strength of the IGMF. The best-fit value is found at $B_0 = 2.8\times 10^{-16}$\,G with a test statistic $\text{TS}=14.2$ when we consider a exponential cutoff in the blazar TeV spectra at $E_{c} = 5$\,TeV, and with $\text{TS}=19.5$ if we assume $E_{c} = 10$\,TeV. }% hypothesis as a function of the strength of IGMF. The TS reaches its maximum value of $23.97$ at $B_0 = 2.3\times 10^{-16}$\,G.   }
%\vspace{-0.5cm}
\label{TS}
\end{figure}

\rev{The intrinsic spectral parameters $\alpha$ and $E_c$ and the field strength $B_0$ are not independent inputs in general. For a given dataset, different assumed magnetic field configurations can favor different intrinsic source models \cite{Batista_2020}. We note that in that work the inferred spectral index is essentially insensitive to the magnetic field, which motivates keeping $\alpha$ fixed to the 4FGL-DR4 value. The relevant degeneracy is therefore between $B_0$ and the cutoff energy $E_c$. To assess the impact of the assumed cutoff, we treat $E_c$ as a nuisance parameter and profile the likelihood over it (Figure \ref{TS}). The best-fit field strength remains at $B_0 = 2.8\times10^{-16}$~G for $E_c$ of 5 and 10\,TeV. This partial decoupling has a physical origin specific to our method: $\alpha$ and $E_c$ set the normalization of the secondary emission, whereas $B_0$ is constrained by the spatial extension of the cascade and, in particular, its asymmetry along the projected jet axis.} We assumed that the intrinsic spectrum of our targets follows a power law spectrum $dN/dE\propto E^{-\alpha}\ \text{exp} \left(-{E}/{E_c}\right)$. 
%The spectral index $\alpha$ is retrieved from the Fermi Point Source Catalog \citep[4FGL-DR4,][]{2022ApJS..263...24A}. We used two different energy cutoffs, $E_c = 5\, \text{TeV}$ and $E_c = 10\, \text{TeV}$ for all of our targets.
The spectral index $\alpha$ is taken from the \textit{Fermi}-LAT Point Source Catalog \cite[4FGL-DR4,][]{2022ApJS..263...24A}, which we assumed is a good representation of the intrinsic spectral slope before EBL absorption. We adopt two representative cutoff energies, $E_c = 5,\mathrm{TeV}$ and $E_c = 10,\mathrm{TeV}$, for all sources. The lower value reflects a conservative scenario, motivated by measurements of the intrinsic spectrum of Mrk~421, which indicate a cutoff around $E_c \sim 5$\,TeV \cite{Albert_2022}. In contrast, Mrk~501 shows either no significant cutoff \cite{Albert_2022} or one at $E_c \sim 9.5$\,TeV \cite{2025icrc.confE.883X}, while 1ES~0229+200 shows no evidence for a cutoff up to $\sim15$\,TeV \cite{2007AA...475L...9A}.

\rev{The main setup of CRPropa is as follows.} TeV photons propagate in %Starting from the source, we place 
a turbulent and random magnetic field with a coherence length of $1$\,Mpc and \rev{a Kolmogorov power spectrum} in a uniform cube consisting of $500^3$ cells, each measuring $0.5$\,Mpc. %To speed up the simulation, CRPropa adopts a parametrized thinning algorithm $\eta$ ranging from 0 to 1. 
% Whenever a cascade process occurs, the code draws a random number r, and only when $r<(1-f)^{\eta}$, where f is the fractional energy remaining by the secondary particle, will the particle be traced. 
%For simplicity, we set this thinning parameter to zero, meaning that all the particles are traced. 
For each source in our sample we simulated $20,000$ primary photons \rev{from $30\,\text{GeV}$ to $20\, \text{TeV}$, and traced the arrival positions as they hit the observer at D equal to the distance of a source at redshift $z$.}  %to minimize the effect of %, in which the 
%statistical fluctuations %will have minor effect on our 
%in the calculation of $f^\text{halo}$, which is 
We derived $f^\text{halo} =  N_\text{cascade}/N_\text{total}$ as the ratio between the number secondary photons and the total number photons that reach the observer within an angular distance of $0.4^{\circ}$ from the source location. %,   
%The fractional component can be calculated as 
%$f^\text{halo} = N_\text{cascade}/N_\text{total}$. 
%Uncertainties can also arise from our assumptions of blazars' spectrum. However, we think this calculation is still a fair approximation to the $f^\text{halo}$ parameter as the CRPropa traces all the particles. For each source, 
We calculated $f^\text{halo}$ values for each source %we simulated the particles 
under magnetic field intensities spanning between $10^{-17}$\,G and $10^{-14}$\,G. %, which is the same range we explore using our Monte-Carlo code. 
The $f^\text{halo}$ values for each source and energy bin derived under an IGMF intensity $B_0=2.8 \times 10^{-16}\,\text{G}$ are listed in Table~\ref{bl_list1}. 
%We count the number of cascade gamma rays and the number of primary gamma rays within a $0.4^{\circ}$ radial separation from the source, and within the two energy bins of $30-100$ GeV and $100-300$ GeV. 

With our tabulated values for $f^{\text{halo}}_{ij}$, we conducted the likelihood ratio test for IGMF magnetic field values %that spans %the magnetic field 
in the range 
$10^{-17}\,\text{G} \leq B_0 \leq 10^{-14}\,\text{G}$ sampled in 32 steps. The likelihood ratio of a pair halo hypothesis ($H_1$) against a null hypothesis ($H_0$) where the magnetic field is zero and no pair halo secondary photons would be observed can be written as
\begin{equation}
    \Lambda(x|D) = \frac{\text{sup}\{L(x,H_0|D)\}}{\text{sup}\{L(x,H_1|D)\}} 
\end{equation}
where ``sup'' is the supremum function. 
The corresponding test statistic is then calculated as $\text{TS} = -2\ \text{ln}\Lambda$. The results of our scan of the Likelihood function as a function of the IGMF magnetic field are shown in Figure \ref{TS}.
%
% Table \ref{bl_list1} is a list of TS value at $B_0 = 2.3\times 10^{-16}$ G for individual sources.
%TODO: add uncertainty: likelihood ratio test vs confidence integral
%TODO: Confidence Level from likelihood ratio test. DOUBLE CHECK!!
%
The only degree of freedom in our model is $B_0$, % TS analysis has one degree of freedom which is 
the strength of IGMF. Therefore, the confidence interval for a likelihood ratio test is given by
%\begin{equation}
$    \text{TS} = -2\ \ln{\Lambda< \chi^2_{1,\alpha}} $
%\end{equation}
where $\chi^2_{1,\alpha}$ indicates the chi-squared distribution with 1 degree of freedom at a 
%$100(1-\alpha)\%$ 
$1-\alpha$
confidence level. %At a $68\%$ confidence level, the $\chi^2_{1,0.32}\approx 1.148$. 

In our more conservative scenario where the TeV spectra of the selected HBLs show an exponential cutoff at $E_c=5$\,TeV we report a hint of a pair halo signal in the GeV \lat data that is compatible with an IGMF with $B_0 = 2.8\times 10^{-16}$\,G. The null hypothesis in which no pair halo signal is present is rejected at $3.8\,\sigma$ confidence level, and 
the $99\%$ confidence interval for the IGMF intensity resulting from our Likelihood function profile is $0.9\times 10^{-16}\ \text{G}<B_0<8.9\times10^{-16}\ \text{G}$. 

If we assume instead that the intrinsic cutoff energy in the TeV blazar spectra is at $E_c=10$\,TeV, the best-fit to our data still appears to be at $B_0 = 2.8\times 10^{-16}$\,G, but the significance of the signal increases to $4.4\,\sigma$ (Figure~\ref{TS}). % We report a stronger signal: $\sim4.4\sigma$, assuming a higher energy cutoff $E_c = 10\,\text{TeV}$, which suggests that the data prefers a larger fraction of secondary gamma rays. 

\section{Discussion and implications of a $10^{-16}$\,G IGMF}
%TODO: use realtilistic spectrum. 
\rev{Assuming a conservative intrinsic cutoff energy of $E_c = 5\text{ TeV}$}, we find evidence for a non-zero intergalactic magnetic field at the level of 
$B_0 \sim 3\times 10^{-16}$\,G, with the null hypothesis excluded at a confidence level of $3.8\sigma$. \rev{Assuming a higher intrinsic cutoff energy of $E_c = 10\text{ TeV}$, the significance increases to $4.4\,\sigma$}. This result strengthens the case for magnetized cosmic voids and demonstrates the power of anisotropic pair-halo searches as a probe of the intergalactic magnetic field.

The magnetic field strength inferred from our analysis, $B_0 \sim 10^{-16}\,\mathrm{G}$, has important implications for the origin and structure of intergalactic magnetic fields. At this level, the field is strong enough to induce observable deflections of cascade pairs over cosmological distances, while remaining well below current upper limits of $B_0 \lesssim 10^{-8}$\,G derived from Faraday rotation measurements \cite{2015MNRAS.452.2851P,2019ApJ...878...92V,2019AA...622A..16O,2020AA...638A..48S,2021MNRAS.503.2913A}. This inferred field strength is consistent with previous searches for extended secondary GeV emission from TeV blazars \cite{Hess_2014,Chen_2015,Archambault_2017}, as well as constraints derived from time-delayed secondary emission (pair echoes) from gamma-ray bursts \cite{wang_2020,huang_2023,mirabal_2023,vovk_2024}, and is broadly compatible with limits obtained from observations of 1ES~0229+200 assuming a coherence length $\lambda > 0.2\,\mathrm{Mpc}$ \cite{Magic_2023}. 
%However, some studies based on spectral signatures of cascade emission have derived stronger lower bounds on the IGMF, with $B \geq 3\times 10^{-16}$\,G \cite{Neronov_2010} and even $B \geq 7.1\times10^{-16}$\,G \cite{Aharonian_2023}, in tension with our result.

However, our result is in tension with recent constraints derived from joint \textit{Fermi}-LAT and H.E.S.S. observations \cite{Aharonian_2023}, which place a lower limit of $B_0 > 7.1\times10^{-16}$\,G at 95\% confidence level for a coherence length of 1~Mpc, even under the assumption of short blazar activity times. This lower limit is higher than our best-fit value of $B_0 \sim 10^{-16}\,\mathrm{G}$. 
We note that the two analyses probe different observational signatures of the cascade emission. The sensitivity of our method comes from the anisotropic spatial distribution of pair halos, which align with the projected jet direction on the sky \cite{Neronov_2010_2}. In contrast, \cite{Aharonian_2023} assumes a viewing angle of $\theta_j=0^\circ$, corresponding to observations directly along the jet axis, which effectively suppresses any asymmetry in the modeled pair halo. This fundamental difference in assumptions may account for the apparent tension between our results.

An IGMF with intensity $B_0 \sim 10^{-16}\,\mathrm{G}$ is broadly consistent with scenarios in which weak seed fields are generated in the early Universe and subsequently evolve without significant amplification in cosmic voids. In contrast, models in which the IGMF is primarily generated by astrophysical processes, such as galactic outflows or large-scale jets, must account for both the apparent large filling factor and the coherence properties required to produce the anisotropic cascade signatures probed in this work. Our result therefore favors a picture in which intergalactic magnetic fields are widespread and dynamically relevant on Mpc scales, although their precise origin remains uncertain. 

Recent studies have shown that strong first-order phase transitions in the early Universe can simultaneously generate primordial magnetic fields with strengths compatible with blazar observations and stochastic gravitational wave backgrounds, providing a unified cosmological origin for the IGMF \cite{borah2026}. In particular, axion-like particle frameworks can produce magnetic fields with amplitudes and coherence lengths consistent with current observational constraints, highlighting the potential connection between early-Universe symmetry breaking and intergalactic magnetism.
%The origin of the IGMF is still an open question. %There are discussions on the origin of IGMF: whether the ``seed'' field is primordial or astrophysical. 
A primordial origin has also been favored in %by Tjemsland, Meyer \& Vazza 
\cite{Tjemsland_2024}, arguing that  the absence of secondary GeV photons in the spectrum of 1ES~0229+200 requires a %is favored because of the 
large fill factor for the IGMF, %of %a large fraction of space 
%$f>0.67$ %filled by a IGMF 
disfavoring scenarios that propose an astrophysical origin. % scenarios \cite{Tjemsland_2024}. 
%Modifications of IGMF from powerful jets are also discussed in one recent study in which they discovered an approximately $7$ Mpc jet. This unexpected long jet occupies $66\%$ of cosmic radius of nearby void, and it is suggested that such an extended jet could potentially alter the magnetic field between the galaxies \cite{Oei_2024}. 

At the same time, the recent discovery on a $\sim7$\,Mpc radio jet \cite{Oei_2024} %that occupies 66\% of the radius of a typical cosmic void has raised the possibility that such extreme structures could influence the magnetic field between galaxies. However, the prevalence and distribution of jets at this scale remain poorly constrained, preventing a robust assessment of their potential impact on the strength of the IGMF on cosmic scales. %, further investigation into the abundance and distribution of such extended jets is necessary.
suggests that large-scale astrophysical processes may also contribute to magnetizing the intergalactic medium \cite{2013AARv..21...62D,2017CQGra..34w4001V}. However, the prevalence and spatial distribution of such spatially extended jets remain poorly constrained, preventing a robust assessment of their overall contribution to the IGMF on cosmological scales. 

% Their analysis is not affected by the possible suppression of pair cascades due to pairs losing energy in the intergalactic medium though plasma instabilities, because the blazar activity timescale assumed in the paper is short $\sim 10$ years compared to the plasma instability timescale \cite{Broderick_2012}. Another paper investigating the same sources has given a more conservative lower limit, $B>2\times 10^{-17}$ G \cite{blunier2025}. However, the assumptions with respect to the blazar activity timescale can have significant impact on the lower bounds that they derived using spectral analysis. To search for the spatial extension of secondary gamma rays, we have assumed that the blazar activity timescale is very large $\sim10^7-10^8$ years. For such large timscale, plasma instabilities might develop and cool the pairs before they up-scatter CMB photons \cite{Broderick_2012,Schlickeiser_2013}. However, growing evidence has ruled out the possibilities that the loss of pairs' energy through plasma instabilities \cite{Arrowsmith_2025, Sironi_2014,Rafighi_2017,Perry_2021}. The discrepancy between our result and the lower limit derived in Aharonian et al. \cite{Aharonian_2023} worsen as the timescale becomes $10^{7}$ yr: $B>3.9\times10^{-14}$ G. This can be studied with better instrumentation. 

An important source of uncertainty in the interpretation of cascade-based constraints is the potential suppression of secondary emission as a result of plasma instabilities in the intergalactic medium. If such instabilities efficiently dissipate the energy of electron-positron pairs before inverse Compton scattering, the observable GeV signal could be significantly reduced. Early studies suggested that this effect could weaken or invalidate cascade-based limits on the IGMF \cite{Broderick_2012, Schlickeiser_2013}. However, more recent work disfavors efficient energy loss through this channel, arguing that plasma instabilities are unlikely to dominate over Compton cooling \cite{miniati2013, sironi2014, Rafighi_2017,Perry_2021, Arrowsmith_2025}. While this remains an active area of research, current evidence supports the interpretation of cascade emission as a robust probe of intergalactic magnetic fields.

Our analysis assumes that TeV blazars are active over long timescales ($\sim 10^7$--$10^8$~yr), allowing pair cascades to fully develop and produce an extended halo signal. Shorter activity timescales would reduce the expected angular extent and flux of the cascade emission, potentially affecting the inferred IGMF constraints \cite{2011ApJ...733L..21D}. This dependence on source history represents an important systematic uncertainty that is common to all cascade-based methods. \rev{In our analysis, we adopt long-term, time-averaged spectra from 4FGL-DR4 to model the primary blazar emission. High-energy variability in sources such as Mrk 421 and Mrk 180 has minimal impact on our search for extended pair halos, as the secondary cascade emission represents a time-integrated superposition of the primary TeV flux over extended propagation timescales, $t\sim10^{6}$. Any short term flares were smoothed out by this propagation delay, and therefore the long-term average spectrum is the physically appropriate input for cascade modeling. }

Our results demonstrate that anisotropic pair-halo searches provide a powerful probe of intergalactic magnetic fields, complementary to other indirect methods. By aligning sources according to their jet position angles, we enhance sensitivity to extended emission components that would otherwise be diluted in isotropic stacking analyses. With future improvements in angular resolution and source characterization, this technique has the potential to transition from statistical detection to direct imaging, opening a new observational window on magnetism in cosmic voids and its origin in the early Universe.

\section{Prospects of detection of the IGMF with future gamma-ray instruments}

Looking ahead, future gamma-ray instruments will play a critical role in improving the sensitivity of pair halo searches. In the following, we estimate what observations of a single source with future gamma-ray instruments will be able to resolve haloes on individual sources, under the assumption that the IGMF takes an intensity equal to our best-fit value of $B_0 =2.8\times 10^{-16}$\,G. 
 
%However, its angular resolution at $\sim 100$~GeV remains insufficient to directly resolve the extended pair halo structures predicted in this work. In contrast, instruments with improved angular resolution in the GeV range, such as GRAINE, may offer the opportunity to directly image anisotropic pair halos and test the predictions of cascade models. 

\begin{figure}[tb]
\centering
\includegraphics[width=.48\linewidth]{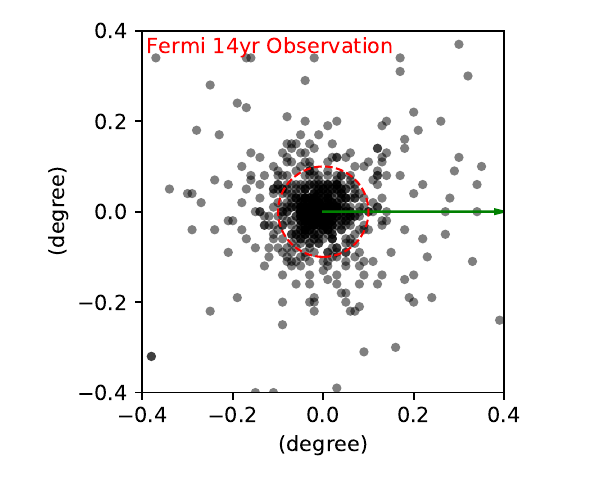}
\includegraphics[width=.48\linewidth]{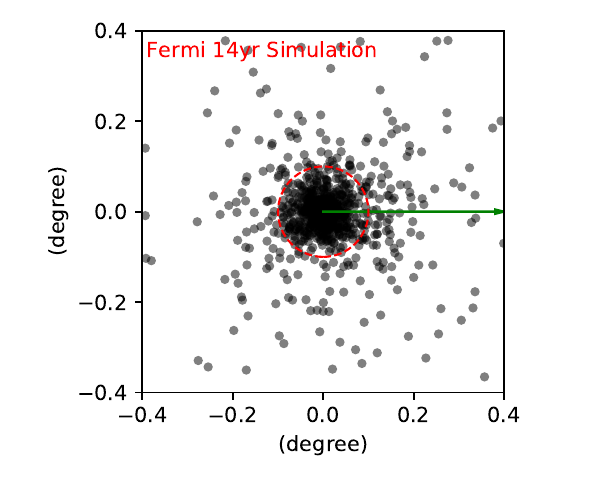}
\caption{\textit{Left:} \lat sky map of Mrk~421 in the 30-100\,GeV energy range.  %The figure on the left is the 14-years Fermi-LAT sky map of Mrk 421 in the 30-100 GeV energy bin. 
The green arrow shows the projected direction of the jet in the sky and the red circle shows the LAT $68\%$ containment circle. \textit{Right: }Simulated \lat sky map assuming $B_0 = 2.8\times 10^{-16}$ G, $f^\text{halo} = 0.3$, $\theta_\text{j} = 1^\circ$ for secondary gamma rays with energy $50$\,GeV.  }
%\vspace{-0.3cm}
\label{fermi_mrk421}
\end{figure}

\begin{figure}[tb]
\centering
\includegraphics[width=.55\linewidth]{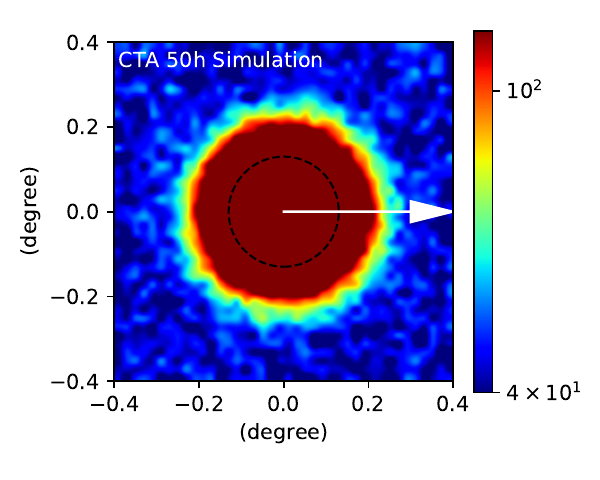}
\caption{Simulated sky maps for 50\,h of observation of Mrk~421 with CTAO % 50h simulated sky maps 
assuming $B_0 = 2.8\times 10^{-16}$ G, $\theta_j = 1^\circ$, $f^\text{halo} = 0.2$. The white arrow shows the projected direction of the jet in the sky and the dark circle shows the CTAO $68\%$ containment circle. 
%TS peaks around 21, which suggest a $4.6\,\sigma$ of pair halo signal.
}
%\vspace{-0.5cm}
\label{ctao_mrk421}
\end{figure}

The Cherenkov Telescope Array Observatory (CTAO) will provide a substantial increase in photon statistics due to its large effective area, enabling more precise measurements of high-energy emission from TeV blazars. The northern array of CTAO has a much larger collection area compared to \lat, improving teh photon statistics for detection of a pair halo. To demonstrate the potential of CTAO compared to \lat, we ran simulations of the bright, nearby blazar Mrk~421. A simulated sky map of Mrk~421 observed by \lat is shown in Figure~\ref{fermi_mrk421}. %The Fermi-LAT has much less observed $\gamma$ ray events at higher energy bin and therefore, is not shown here. 
For CTAO, we simulated the direct and secondary emission that would be observed at energy of $100$\,GeV (Figure~\ref{ctao_mrk421}). %  and with this simulated map, we calculated its test statistics using our pair halo model. 
We assumed a collection area for CTAO of $8\times 10^{4}\,\text{m}^2$, with a 68\% containment angle of $0.13^\circ$ at 100\,GeV \cite{ctao_performance}. We calculated the expected number of counts during 50\,h of observation, using an intrinsic source spectrum from \cite{Abdo_2011}. 
% \begin{equation}
%     \frac{dN}{dE} = N_0 \left(\frac{E}{E_0}\right)^{-\alpha}\times e^{\left(-\frac{E}{E_c}\right)}\times e^{-\tau(E,z)}
% \end{equation}
% where normalization flux is given as $N_0 = 4.0\times10^{-11}\ \text{TeV}^{-1}\text{cm}^{-2}\text{s}^{-1}$, spectral index is $\alpha = 2.26$, and energy cutoff is $E_c = 5.1\ \text{TeV}$ \cite{Albert_2022}. The optical depth $\tau$ was incorporated in our simulation of pair production process. 
The background emission is assumed to be $0.6\ \text{Hz}\,\text{deg}^{-2}$. 

% Gamma-Ray Astro-Imager with Nuclear Emulsion
Despite the improved sensitivity of CTAO, its angular resolution comparable to that of \lat at $100$\,GeV makes it challenging for CTAO to resolve and characterize individual pair haloes. % , pair halos can not be directly seen from CTAO even with much higher events. 

One detection technology has intrinsically a much better angular resolution than the silicon tracker technology of \lat or the Air Cherenkov technique used by CTAO is the use of nuclear emulsions. In particular, the Gamma-Ray Astro-Imager with Nuclear Emulsion telescope (GRAINE)
%The GRAINE emulsion $\gamma$-ray telescope 
achieves a significantly improved angular resolution compared to current instruments. GRAINE has a design performance reaching $\sim 0.1^\circ$ (5 arcmin) at 1--2 GeV \cite{2024ApJ...960...47T}, corresponding to nearly an order-of-magnitude improvement over the $\sim 0.8^\circ$ resolution of \textit{Fermi}-LAT at similar energies. This gain is enabled by the use of nuclear emulsions, which provide an intrinsic spatial resolution of $\lesssim1\,\mu$m for charged particle tracks, allowing precise reconstruction of the electron--positron pair production vertex with minimal impact from multiple Coulomb scattering. In contrast, silicon tracker detectors such as \textit{Fermi}-LAT have spatial resolutions at the level of tens of microns and rely on thicker converter materials, leading to worse direction reconstruction due tio multiple Coulomb scattering. %larger scattering-induced uncertainties. 
As a result, emulsion-based detectors can reconstruct pair kinematics and arrival directions with substantially higher precision, particularly in the GeV energy range.

\begin{figure}[tb]
\centering
\includegraphics[width=.55\linewidth]{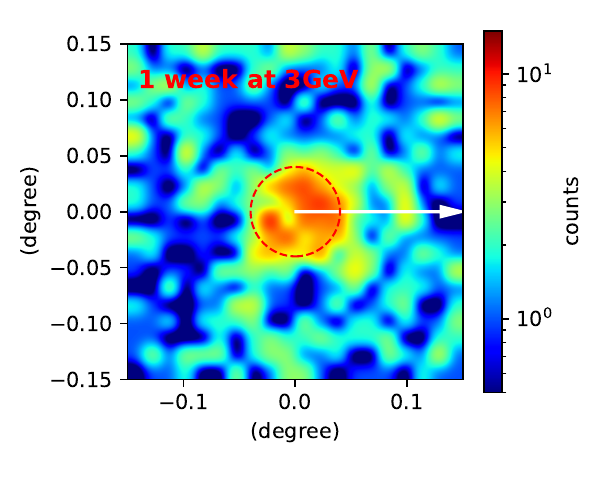}
\caption{Simulated sky map at 3\,GeV for 56\,h  of observation of Mrk~421 with GRAINE observation, corresponding to a 1 week duration flight from Kiruna, Sweden. % 50h simulated sky maps 
We assume $B_0 = 2.8\times 10^{-16}$\,G, $\theta_j = 1^\circ$, $f^\text{halo} = 0.3$. The background is assumed to be $60\,\text{m}^2\text{s}^{-1}\text{sr}^{-1}$. The white arrow shows the projected direction of the jet in the sky and the red circle shows the GRAINE $68\%$ containment circle for comparison.}
%\vspace{-0.5cm}
\label{GRAINE_mrk421}
\end{figure}

We simulate an observation of Mrk~421 with GRAINE. We use a %However, another instrument GRAINE has significantly better angular resolution; 
68\% containment radius of $0.04^\circ$ at $3$\,GeV \cite{2024ApJ...960...47T}. Because nuclear emulsions have to be recovered and scanned to perform the event reconstruction, GRAINE can only fly from stratospheric ballooning platform.  Assuming a 1~week duration ballon flight from the northern hemisphere (Kiruna, Sweden), Mrk~421 could be observed at favorable zenith angles for $\sim8\,h$ per day for 7 days. The effective area of GRAINE is assumed to be $2\,\text{m}^2$. The resulting simulated skymap is shown in Figure~\ref{GRAINE_mrk421}. A hint of asymmetry due to the pair halo signal could be seen in a flight of 1~week duration, and a stronger detection could be achieved by stacking teh signals from successive flights.  

% \begin{figure}[h!]
% \centering
% \includegraphics[width=.45\linewidth]{GraineTS_sim.png}
% \caption{}
% \label{GRAINE_mrk421_TS}
% \end{figure}

%Our analysis would also improve with more accurate measurements of the HBLs' position angles, and more radio observations of HBL sources. 

In summary, while CTAO will significantly improve photon statistics, resolving individual pair halos will ultimately require advances in angular resolution. Improvements in direction reconstruction for air Cherenkov telescope arrays, together with next-generation detector technologies such as nuclear emulsions (GRAINE), may finally enable direct imaging of pair halos around TeV blazars. In contrast, continued observations with \textit{Fermi}-LAT are expected to improve sensitivity only as the squared root of the accumulated exposure, making substantial gains increasingly difficult. However, additional radio interferometry measurements of jet position angles for a larger population of HBLs could expand the number of sources suitable for anisotropic stacking analyses with \lat. This increase in sample size may provide a path toward robust, statistically significant evidence for pair halos in the GeV band.

\newpage

\appendix
% \section{Test Statistics for Individual Source}
% \begin{table}[h!]
% \centering
% \begin{tabular}{c c}
% \hline
% Blazar name & TS \\
% \hline
% IZw 187 & 1.423\\
% Mrk 421 & 7.532\\
% Mrk 180 & -0.0087\\
% 1ES 2344+514 & 0.752\\
% 1ES 1741+196 &  0.649\\
% TXS 0210+515 & 0.750\\
% TXS 0518+211 & -2.508\\
% TXS 1515-273 & 0.453\\
% 1ES 0806+524 & 1.433\\
% 1H 1914-194 &  0.054\\
% PKS 0548-322 & 0.771\\
% RBS 0030 & 0.060\\
% PMN J0152+0146 & -0.111\\
% 1ES 1959+650 & 4.179\\
% 1ES 0229+200 & -0.079\\
% 1H 0658+595 & 0.615\\
% 1RXS J101015.9-311909 & 0.140\\
% H 1426+428 & 0.344\\
% PKS 2155-304 & 4.705\\
% B3 2247+381 & 1.624\\
% PKS 2005-489 & 0.551\\
% \hline
% \end{tabular}
% \caption{A list of TS value for individual sources at $B_0 = 1.81\times 10^{-16}$ G.
% \label{bl_TS_list}}
% \end{table}

\acknowledgments

We thank Bhupal Dev, Francesc Ferrer, and Anastasia Sokolenko for helpful discussions that prompted revisions and improved the quality of this work. We also thank Hiroki Rokujo for sharing the GRAINE response functions. This work was supported by NASA Fermi Guest Observer grant 80NSSC25K7110. 
Authors acknowledge the Fermi-LAT team and VLBA team for 
providing the data that made this study possible. This research has also used the CTAO instrument response function obtained from \url{https://www.ctao.org/for-scientists/performance/}.

% \paragraph{Note added.} This is also a good position for notes added
% after the paper has been written.

% Bibliography

%% [A] Recommended: using JHEP.bst file
\bibliographystyle{JHEP}
\bibliography{reference}

\providecommand{\href}[2]{#2}\begingroup\raggedright\begin{thebibliography}{10}

\bibitem{1999ARAA..37...37K}
R.M.~{Kulsrud}, \emph{{A Critical Review of Galactic Dynamos}}, \href{https://doi.org/10.1146/annurev.astro.37.1.37}{\emph{Ann. Rev. Astron. Atrophys.} {\bfseries 37} (1999) 37}.

\bibitem{Carilli_2002}
C.L.~{Carilli} and G.B.~{Taylor}, \emph{{Cluster Magnetic Fields}}, \href{https://doi.org/10.1146/annurev.astro.40.060401.093852}{\emph{Annual Review of Astronomy and Astrophysics} {\bfseries 40} (2002) 319} [\href{https://arxiv.org/abs/astro-ph/0110655}{{\ttfamily astro-ph/0110655}}].

\bibitem{1950ZNatA...5...65B}
L.~{Biermann}, \emph{{{\"U}ber den Ursprung der Magnetfelder auf Sternen und im interstellaren Raum (miteinem Anhang von A. Schl{\"u}ter)}}, {\emph{Zeitschrift Naturforschung Teil A} {\bfseries 5} (1950) 65}.

\bibitem{Kulsrud_2008}
R.M.~Kulsrud and E.G.~Zweibel, \emph{On the origin of cosmic magnetic fields}, \href{https://doi.org/10.1088/0034-4885/71/4/046901}{\emph{Reports on Progress in Physics} {\bfseries 71} (2008) 046901}.

\bibitem{Grasso_2001}
D.~Grasso and H.R.~Rubinstein, \emph{Magnetic fields in the early universe}, \href{https://doi.org/https://doi.org/10.1016/S0370-1573(00)00110-1}{\emph{Physics Reports} {\bfseries 348} (2001) 163}.

\bibitem{2002RvMP...74..775W}
L.M.~{Widrow}, \emph{{Origin of galactic and extragalactic magnetic fields}}, \href{https://doi.org/10.1103/RevModPhys.74.775}{\emph{Reviews of Modern Physics} {\bfseries 74} (2002) 775} [\href{https://arxiv.org/abs/astro-ph/0207240}{{\ttfamily astro-ph/0207240}}].

\bibitem{Vachaspati_2021}
T.~Vachaspati, \emph{Progress on cosmological magnetic fields}, \href{https://doi.org/10.1088/1361-6633/ac03a9}{\emph{Reports on Progress in Physics} {\bfseries 84} (2021) 074901}.

\bibitem{Planck_2016}
{Planck Collaboration}, {Ade, P. A. R.}, {Aghanim, N.}, {Arnaud, M.}, {Arroja, F.}, {Ashdown, M.} et~al., \emph{Planck 2015 results - xix. constraints on primordial magnetic fields}, \href{https://doi.org/10.1051/0004-6361/201525821}{\emph{A\&A} {\bfseries 594} (2016) A19}.

\bibitem{Pshirkov_2016}
M.S.~Pshirkov, P.G.~Tinyakov and F.R.~Urban, \emph{New limits on extragalactic magnetic fields from rotation measures}, \href{https://doi.org/10.1103/physrevlett.116.191302}{\emph{Physical Review Letters} {\bfseries 116} (2016) }.

\bibitem{Neronov_2009}
A.~Neronov and D.V.~Semikoz, \emph{Sensitivity of $\ensuremath{\gamma}$-ray telescopes for detection of magnetic fields in the intergalactic medium}, \href{https://doi.org/10.1103/PhysRevD.80.123012}{\emph{Phys. Rev. D} {\bfseries 80} (2009) 123012}.

\bibitem{1994ApJ...423L...5A}
F.A.~{Aharonian}, P.S.~{Coppi} and H.J.~{Voelk}, \emph{{Very High Energy Gamma Rays from Active Galactic Nuclei: Cascading on the Cosmic Background Radiation Fields and the Formation of Pair Halos}}, \href{https://doi.org/10.1086/187222}{\emph{\apjl} {\bfseries 423} (1994) L5} [\href{https://arxiv.org/abs/astro-ph/9312045}{{\ttfamily astro-ph/9312045}}].

\bibitem{Domnguez_2010}
A.~Dom{\'\i}nguez, J.R.~Primack, D.J.~Rosario, F.~Prada, R.C.~Gilmore, S.M.~Faber et~al., \emph{Extragalactic background light inferred from aegis galaxy-sed-type fractions: Ebl from aegis galaxy-sed-type fractions}, \href{https://doi.org/10.1111/j.1365-2966.2010.17631.x}{\emph{Monthly Notices of the Royal Astronomical Society} {\bfseries 410} (2010) 2556}.

\bibitem{chen_2018}
W.~{Chen}, M.~{Errando} and J.~{Buckley}, \emph{{Novel search for TeV-initiated pair cascades in the intergalactic medium}},  in \emph{42nd COSPAR Scientific Assembly}, vol.~42, pp.~E1.14--26--18, July, 2018, \href{https://doi.org/10.48550/arXiv.1811.05774}{DOI} [\href{https://arxiv.org/abs/1811.05774}{{\ttfamily 1811.05774}}].

\bibitem{Long_2015}
A.J.~Long and T.~Vachaspati, \emph{Morphology of blazar-induced gamma ray halos due to a helical intergalactic magnetic field}, \href{https://doi.org/10.1088/1475-7516/2015/09/065}{\emph{Journal of Cosmology and Astroparticle Physics} {\bfseries 2015} (2015) 065}.

\bibitem{2016ApJ...832..109B}
A.E.~{Broderick}, P.~{Tiede}, M.~{Shalaby}, C.~{Pfrommer}, E.~{Puchwein}, P.~{Chang} et~al., \emph{{Bow Ties in the Sky. I: The Angular Structure of Inverse Compton Gamma-Ray Halos in the Fermi Sky}}, \href{https://doi.org/10.3847/0004-637X/832/2/109}{\emph{\apj} {\bfseries 832} (2016) 109} [\href{https://arxiv.org/abs/1609.00387}{{\ttfamily 1609.00387}}].

\bibitem{Batista}
R.~Alves~Batista, A.~Saveliev, G.~Sigl and T.~Vachaspati, \emph{Probing intergalactic magnetic fields with simulations of electromagnetic cascades}, \href{https://doi.org/10.1103/PhysRevD.94.083005}{\emph{Phys. Rev. D} {\bfseries 94} (2016) 083005}.

\bibitem{Neronov_2010}
A.~Neronov and I.~Vovk, \emph{Evidence for strong extragalactic magnetic fields from fermi observations of tev blazars}, \href{https://doi.org/10.1126/science.1184192}{\emph{Science} {\bfseries 328} (2010) 73} [\href{https://arxiv.org/abs/https://www.science.org/doi/pdf/10.1126/science.1184192}{{\ttfamily https://www.science.org/doi/pdf/10.1126/science.1184192}}].

\bibitem{Chen_2015}
W.~Chen, J.H.~Buckley and F.~Ferrer, \emph{Search for gev $\ensuremath{\gamma}$-ray pair halos around low redshift blazars}, \href{https://doi.org/10.1103/PhysRevLett.115.211103}{\emph{Phys. Rev. Lett.} {\bfseries 115} (2015) 211103}.

\bibitem{Aleksic_2010}
J.~{Aleksi{\'c}}, L.A.~{Antonelli}, P.~{Antoranz}, M.~{Backes}, C.~{Baixeras}, J.A.~{Barrio} et~al., \emph{{Search for an extended VHE {\ensuremath{\gamma}}-ray emission from Mrk 421 and Mrk 501 with the MAGIC Telescope}}, \href{https://doi.org/10.1051/0004-6361/201014747}{\emph{\aap} {\bfseries 524} (2010) A77} [\href{https://arxiv.org/abs/1004.1093}{{\ttfamily 1004.1093}}].

\bibitem{Hess_2014}
{H.E.S.S. Collaboration:}, {Abramowski, A.}, {Aharonian, F.}, {Ait Benkhali, F.}, {Akhperjanian, A. G.}, {Ang{\"u}ner, E.} et~al., \emph{Search for extended $\ensuremath{\gamma}$-ray emission around agn with h.e.s.s. and fermi-lat}, \href{https://doi.org/10.1051/0004-6361/201322510}{\emph{A\&A} {\bfseries 562} (2014) A145}.

\bibitem{Archambault_2017}
S.~Archambault, A.~Archer, W.~Benbow, M.~Buchovecky, V.~Bugaev, M.~Cerruti et~al., \emph{Search for magnetically broadened cascade emission from blazars with veritas}, \href{https://doi.org/10.3847/1538-4357/835/2/288}{\emph{The Astrophysical Journal} {\bfseries 835} (2017) 288}.

\bibitem{wang_2020}
Z.-R.~{Wang}, S.-Q.~{Xi}, R.-Y.~{Liu}, R.~{Xue} and X.-Y.~{Wang}, \emph{{Constraints on the intergalactic magnetic field from {\ensuremath{\gamma}} -ray observations of GRB 190114C}}, \href{https://doi.org/10.1103/PhysRevD.101.083004}{\emph{\prd} {\bfseries 101} (2020) 083004} [\href{https://arxiv.org/abs/2001.01186}{{\ttfamily 2001.01186}}].

\bibitem{Magic_2023}
V.A.~{Acciari}, I.~{Agudo}, T.~{Aniello}, S.~{Ansoldi}, L.A.~{Antonelli}, A.~{Arbet Engels} et~al., \emph{{A lower bound on intergalactic magnetic fields from time variability of 1ES 0229+200 from MAGIC and Fermi/LAT observations}}, \href{https://doi.org/10.1051/0004-6361/202244126}{\emph{\aap} {\bfseries 670} (2023) A145} [\href{https://arxiv.org/abs/2210.03321}{{\ttfamily 2210.03321}}].

\bibitem{mirabal_2023}
N.~{Mirabal}, \emph{{Secondary GeV-TeV emission from ultra-high-energy cosmic rays accelerated by GRB 221009A}}, \href{https://doi.org/10.1093/mnrasl/slac157}{\emph{\mnras} {\bfseries 519} (2023) L85} [\href{https://arxiv.org/abs/2210.14243}{{\ttfamily 2210.14243}}].

\bibitem{huang_2023}
Y.-Y.~{Huang}, C.-Y.~{Dai}, H.-M.~{Zhang}, R.-Y.~{Liu} and X.-Y.~{Wang}, \emph{{Constraints on the Intergalactic Magnetic Field Strength from {\ensuremath{\gamma}}-Ray Observations of GRB 221009A}}, \href{https://doi.org/10.3847/2041-8213/acf66a}{\emph{\apjl} {\bfseries 955} (2023) L10} [\href{https://arxiv.org/abs/2306.05970}{{\ttfamily 2306.05970}}].

\bibitem{vovk_2024}
I.~{Vovk}, A.~{Korochkin}, A.~{Neronov} and D.~{Semikoz}, \emph{{Constraints on the intergalactic magnetic field from Fermi/LAT observations of the `pair echo' of GRB 221009A}}, \href{https://doi.org/10.1051/0004-6361/202347310}{\emph{\aap} {\bfseries 683} (2024) A25}.

\bibitem{2021Univ....7..223A}
R.~{Alves Batista} and A.~{Saveliev}, \emph{{The Gamma-Ray Window to Intergalactic Magnetism}}, \href{https://doi.org/10.3390/universe7070223}{\emph{Universe} {\bfseries 7} (2021) 223} [\href{https://arxiv.org/abs/2105.12020}{{\ttfamily 2105.12020}}].

\bibitem{Pushkarev2017}
A.B.~{Pushkarev}, Y.Y.~{Kovalev}, M.L.~{Lister} and T.~{Savolainen}, \emph{{MOJAVE - XIV. Shapes and opening angles of AGN jets}}, \href{https://doi.org/10.1093/mnras/stx854}{\emph{\mnras} {\bfseries 468} (2017) 4992} [\href{https://arxiv.org/abs/1705.02888}{{\ttfamily 1705.02888}}].

\bibitem{Neronov_2010_2}
A.~Neronov, D.~Semikoz, M.~Kachelriess, S.~Ostapchenko and A.~Elyiv, \emph{Degree-scale gev ``jets'' from active and dead tev blazars}, \href{https://doi.org/10.1088/2041-8205/719/2/l130}{\emph{The Astrophysical Journal} {\bfseries 719} (2010) L130}.

\bibitem{Franceschini2008}
A.~{Franceschini}, G.~{Rodighiero} and M.~{Vaccari}, \emph{{Extragalactic optical-infrared background radiation, its time evolution and the cosmic photon-photon opacity}}, \href{https://doi.org/10.1051/0004-6361:200809691}{\emph{\aap} {\bfseries 487} (2008) 837} [\href{https://arxiv.org/abs/0805.1841}{{\ttfamily 0805.1841}}].

\bibitem{Aharonian_2023}
F.~Aharonian, J.~Aschersleben, M.~Backes, V.B.~Martins, R.~Batzofin, Y.~Becherini et~al., \emph{Constraints on the intergalactic magnetic field using fermi-lat and h.e.s.s. blazar observations}, \href{https://doi.org/10.3847/2041-8213/acd777}{\emph{The Astrophysical Journal Letters} {\bfseries 950} (2023) L16}.

\bibitem{2022ApJS..263...24A}
M.~{Ajello}, L.~{Baldini}, J.~{Ballet}, D.~{Bastieri}, J.~{Becerra Gonzalez}, R.~{Bellazzini} et~al., \emph{{The Fourth Catalog of Active Galactic Nuclei Detected by the Fermi Large Area Telescope: Data Release 3}}, \href{https://doi.org/10.3847/1538-4365/ac9523}{\emph{\apjs} {\bfseries 263} (2022) 24} [\href{https://arxiv.org/abs/2209.12070}{{\ttfamily 2209.12070}}].

\bibitem{2017ApJS..232...18A}
M.~{Ajello}, W.B.~{Atwood}, L.~{Baldini}, J.~{Ballet}, G.~{Barbiellini}, D.~{Bastieri} et~al., \emph{{3FHL: The Third Catalog of Hard Fermi-LAT Sources}}, \href{https://doi.org/10.3847/1538-4365/aa8221}{\emph{\apjs} {\bfseries 232} (2017) 18} [\href{https://arxiv.org/abs/1702.00664}{{\ttfamily 1702.00664}}].

\bibitem{Lister_2019}
M.L.~Lister, D.C.~Homan, T.~Hovatta, K.I.~Kellermann, S.~Kiehlmann, Y.Y.~Kovalev et~al., \emph{Mojave. xvii. jet kinematics and parent population properties of relativistically beamed radio-loud blazars}, \href{https://doi.org/10.3847/1538-4357/ab08ee}{\emph{The Astrophysical Journal} {\bfseries 874} (2019) 43}.

\bibitem{2022ApJS..260...53A}
S.~{Abdollahi}, F.~{Acero}, L.~{Baldini}, J.~{Ballet}, D.~{Bastieri}, R.~{Bellazzini} et~al., \emph{{Incremental Fermi Large Area Telescope Fourth Source Catalog}}, \href{https://doi.org/10.3847/1538-4365/ac6751}{\emph{\apjs} {\bfseries 260} (2022) 53} [\href{https://arxiv.org/abs/2201.11184}{{\ttfamily 2201.11184}}].

\bibitem{piner_2018}
B.G.~{Piner} and P.G.~{Edwards}, \emph{{Multi-epoch VLBA Imaging of 20 New TeV Blazars: Apparent Jet Speeds}}, \href{https://doi.org/10.3847/1538-4357/aaa425}{\emph{\apj} {\bfseries 853} (2018) 68} [\href{https://arxiv.org/abs/1801.00817}{{\ttfamily 1801.00817}}].

\bibitem{TANAMI}
{Ojha, R.}, {Kadler, M.}, {B{\"o}ck, M.}, {Booth, R.}, {Dutka, M. S.}, {Edwards, P. G.} et~al., \emph{Tanami: tracking active galactic nuclei with austral milliarcsecond interferometry - i. first-epoch 8.4 ghz images}, \href{https://doi.org/10.1051/0004-6361/200912724}{\emph{A\&A} {\bfseries 519} (2010) A45}.

\bibitem{Alves_Batista_2022}
R.~Alves~Batista, J.~Becker~Tjus, J.~D{\"o}rner, A.~Dundovic, B.~Eichmann, A.~Frie et~al., \emph{Crpropa 3.2 --- an advanced framework for high-energy particle propagation in extragalactic and galactic spaces}, \href{https://doi.org/10.1088/1475-7516/2022/09/035}{\emph{Journal of Cosmology and Astroparticle Physics} {\bfseries 2022} (2022) 035}.

\bibitem{Batista_2020}
R.A.~Batista and A.~Saveliev, \emph{Multimessenger constraints on intergalactic magnetic fields from the flare of txs 0506+056}, \href{https://doi.org/10.3847/2041-8213/abb816}{\emph{The Astrophysical Journal Letters} {\bfseries 902} (2020) L11}.

\bibitem{Albert_2022}
A.~Albert, R.~Alfaro, C.~Alvarez, J.R.A.~Camacho, J.C.~Arteaga-Vel{\'a}zquez, K.P.~Arunbabu et~al., \emph{Long-term spectra of the blazars mrk 421 and mrk 501 at tev energies seen by hawc}, \href{https://doi.org/10.3847/1538-4357/ac58f6}{\emph{The Astrophysical Journal} {\bfseries 929} (2022) 125}.

\bibitem{2025icrc.confE.883X}
D.~{Xiao}, S.~{Cui}, M.~{Zha}, B.T.~{Zhang}, S.~{Chen} and {Lhaaso Collaboration}, \emph{{Long-term observation of the blazars Mrk 501 by LHAASO}},  in \emph{39th International Cosmic Ray Conference}, p.~883, Dec., 2025.

\bibitem{2007AA...475L...9A}
F.~{Aharonian}, A.G.~{Akhperjanian}, U.~{Barres de Almeida}, A.R.~{Bazer-Bachi}, B.~{Behera}, M.~{Beilicke} et~al., \emph{{New constraints on the mid-IR EBL from the HESS discovery of VHE {\ensuremath{\gamma}}-rays from 1ES 0229+200}}, \href{https://doi.org/10.1051/0004-6361:20078462}{\emph{\aap} {\bfseries 475} (2007) L9} [\href{https://arxiv.org/abs/0709.4584}{{\ttfamily 0709.4584}}].

\bibitem{2015MNRAS.452.2851P}
M.S.~{Pshirkov}, P.G.~{Tinyakov} and F.R.~{Urban}, \emph{{The rotation measures of high-luminosity sources as seen from the NVSS}}, \href{https://doi.org/10.1093/mnras/stv1273}{\emph{\mnras} {\bfseries 452} (2015) 2851} [\href{https://arxiv.org/abs/1407.3909}{{\ttfamily 1407.3909}}].

\bibitem{2019ApJ...878...92V}
T.~{Vernstrom}, B.M.~{Gaensler}, L.~{Rudnick} and H.~{Andernach}, \emph{{Differences in Faraday Rotation between Adjacent Extragalactic Radio Sources as a Probe of Cosmic Magnetic Fields}}, \href{https://doi.org/10.3847/1538-4357/ab1f83}{\emph{\apj} {\bfseries 878} (2019) 92} [\href{https://arxiv.org/abs/1905.02410}{{\ttfamily 1905.02410}}].

\bibitem{2019AA...622A..16O}
S.P.~{O'Sullivan}, J.~{Machalski}, C.L.~{Van Eck}, G.~{Heald}, M.~{Br{\"u}ggen}, J.P.U.~{Fynbo} et~al., \emph{{The intergalactic magnetic field probed by a giant radio galaxy}}, \href{https://doi.org/10.1051/0004-6361/201833832}{\emph{\aap} {\bfseries 622} (2019) A16} [\href{https://arxiv.org/abs/1811.07934}{{\ttfamily 1811.07934}}].

\bibitem{2020AA...638A..48S}
C.~{Stuardi}, S.P.~{O'Sullivan}, A.~{Bonafede}, M.~{Br{\"u}ggen}, P.~{Dabhade}, C.~{Horellou} et~al., \emph{{The LOFAR view of intergalactic magnetic fields with giant radio galaxies}}, \href{https://doi.org/10.1051/0004-6361/202037635}{\emph{\aap} {\bfseries 638} (2020) A48} [\href{https://arxiv.org/abs/2004.05169}{{\ttfamily 2004.05169}}].

\bibitem{2021MNRAS.503.2913A}
A.D.~{Amaral}, T.~{Vernstrom} and B.M.~{Gaensler}, \emph{{Constraints on large-scale magnetic fields in the intergalactic medium using cross-correlation methods}}, \href{https://doi.org/10.1093/mnras/stab564}{\emph{\mnras} {\bfseries 503} (2021) 2913} [\href{https://arxiv.org/abs/2102.11312}{{\ttfamily 2102.11312}}].

\bibitem{borah2026}
P.~{Borah}, P.S.~{Bhupal Dev} and A.~{Ghoshal}, \emph{{Primordial Magnetogenesis and Gravitational Waves from ALP-assisted Phase Transition}}, \href{https://doi.org/10.48550/arXiv.2604.20768}{\emph{arXiv e-prints} (2026) arXiv:2604.20768} [\href{https://arxiv.org/abs/2604.20768}{{\ttfamily 2604.20768}}].

\bibitem{Tjemsland_2024}
J.~Tjemsland, M.~Meyer and F.~Vazza, \emph{Constraining the astrophysical origin of intergalactic magnetic fields}, \href{https://doi.org/10.3847/1538-4357/ad22dd}{\emph{The Astrophysical Journal} {\bfseries 963} (2024) 135}.

\bibitem{Oei_2024}
M.S.S.L.~Oei, M.J.~Hardcastle, R.~Timmerman, A.R.D.J.G.I.B.~Gast, A.~Botteon, A.C.~Rodriguez et~al., \emph{Black hole jets on the scale of the cosmic web}, \href{https://doi.org/10.1038/s41586-024-07879-y}{\emph{Nature} {\bfseries 633} (2024) 537}.

\bibitem{2013AARv..21...62D}
R.~{Durrer} and A.~{Neronov}, \emph{{Cosmological magnetic fields: their generation, evolution and observation}}, \href{https://doi.org/10.1007/s00159-013-0062-7}{\emph{\aapr} {\bfseries 21} (2013) 62} [\href{https://arxiv.org/abs/1303.7121}{{\ttfamily 1303.7121}}].

\bibitem{2017CQGra..34w4001V}
F.~{Vazza}, M.~{Br{\"u}ggen}, C.~{Gheller}, S.~{Hackstein}, D.~{Wittor} and P.M.~{Hinz}, \emph{{Simulations of extragalactic magnetic fields and of their observables}}, \href{https://doi.org/10.1088/1361-6382/aa8e60}{\emph{Classical and Quantum Gravity} {\bfseries 34} (2017) 234001} [\href{https://arxiv.org/abs/1711.02669}{{\ttfamily 1711.02669}}].

\bibitem{Broderick_2012}
A.E.~{Broderick}, P.~{Chang} and C.~{Pfrommer}, \emph{{The Cosmological Impact of Luminous TeV Blazars. I. Implications of Plasma Instabilities for the Intergalactic Magnetic Field and Extragalactic Gamma-Ray Background}}, \href{https://doi.org/10.1088/0004-637X/752/1/22}{\emph{\apj} {\bfseries 752} (2012) 22} [\href{https://arxiv.org/abs/1106.5494}{{\ttfamily 1106.5494}}].

\bibitem{Schlickeiser_2013}
R.~Schlickeiser, S.~Krakau and M.~Supsar, \emph{Plasma effects on fast pair beams. ii. reactive versus kinetic instability of parallel electrostatic waves}, \href{https://doi.org/10.1088/0004-637x/777/1/49}{\emph{The Astrophysical Journal} {\bfseries 777} (2013) 49}.

\bibitem{miniati2013}
F.~{Miniati} and A.~{Elyiv}, \emph{{Relaxation of Blazar-induced Pair Beams in Cosmic Voids}}, \href{https://doi.org/10.1088/0004-637X/770/1/54}{\emph{\apj} {\bfseries 770} (2013) 54} [\href{https://arxiv.org/abs/1208.1761}{{\ttfamily 1208.1761}}].

\bibitem{sironi2014}
L.~{Sironi} and D.~{Giannios}, \emph{{Relativistic Pair Beams from TeV Blazars: A Source of Reprocessed GeV Emission rather than Intergalactic Heating}}, \href{https://doi.org/10.1088/0004-637X/787/1/49}{\emph{\apj} {\bfseries 787} (2014) 49} [\href{https://arxiv.org/abs/1312.4538}{{\ttfamily 1312.4538}}].

\bibitem{Rafighi_2017}
I.~Rafighi, S.~Vafin, M.~Pohl and J.~Niemiec, \emph{Plasma effects on relativistic pair beams from tev blazars: Pic simulations and analytical predictions}, \href{https://doi.org/10.1051/0004-6361/201731127}{\emph{Astronomy \& Astrophysics} {\bfseries 607} (2017) A112}.

\bibitem{Perry_2021}
R.~Perry and Y.~Lyubarsky, \emph{The role of resonant plasma instabilities in the evolution of blazar-induced pair beams}, \href{https://doi.org/10.1093/mnras/stab324}{\emph{Monthly Notices of the Royal Astronomical Society} {\bfseries 503} (2021) 2215}.

\bibitem{Arrowsmith_2025}
C.D.~Arrowsmith, F.~Miniati, P.J.~Bilbao, P.~Simon, A.F.A.~Bott, S.~Burger et~al., \emph{Suppression of pair beam instabilities in a laboratory analogue of blazar pair cascades}, \href{https://doi.org/10.1073/pnas.2513365122}{\emph{Proceedings of the National Academy of Sciences} {\bfseries 122} (2025) }.

\bibitem{2011ApJ...733L..21D}
C.D.~{Dermer}, M.~{Cavadini}, S.~{Razzaque}, J.D.~{Finke}, J.~{Chiang} and B.~{Lott}, \emph{{Time Delay of Cascade Radiation for TeV Blazars and the Measurement of the Intergalactic Magnetic Field}}, \href{https://doi.org/10.1088/2041-8205/733/2/L21}{\emph{\apjl} {\bfseries 733} (2011) L21} [\href{https://arxiv.org/abs/1011.6660}{{\ttfamily 1011.6660}}].

\bibitem{ctao_performance}
``Ctao performance.''

\bibitem{Abdo_2011}
A.A.~Abdo, M.~Ackermann, M.~Ajello, L.~Baldini, J.~Ballet, G.~Barbiellini et~al., \emph{Fermi large area telescope observations of markarian 421: The missing piece of its spectral energy distribution}, \href{https://doi.org/10.1088/0004-637X/736/2/131}{\emph{The Astrophysical Journal} {\bfseries 736} (2011) 131}.

\bibitem{2024ApJ...960...47T}
S.~{Takahashi}, S.~{Aoki}, A.~{Iyono}, A.~{Karasuno}, K.~{Kodama}, R.~{Komatani} et~al., \emph{{First Emulsion {\ensuremath{\gamma}}-Ray Telescope Imaging of the Vela Pulsar by the GRAINE 2018 Balloon-borne Experiment}}, \href{https://doi.org/10.3847/1538-4357/ad0973}{\emph{\apj} {\bfseries 960} (2024) 47}.

\end{thebibliography}\endgroup

%% or
%% [B] Manual formatting (see below)
%% (i) We suggest to always provide author, title and journal data or doi:
%% in short all the informations that clearly identify a document.
%% (ii) please avoid comments such as "For a review'', "For some examples",
%% "and references therein" or move them in the text. In general, please leave only references in the bibliography and move all
%% accessory text in footnotes.
%% (iii) Also, please have only one work for each \bibitem.
% \addbibresource{biblio.bib} 
% \begin{thebibliography}{99}

% % \bibitem{a}
% % Author,
% % \emph{Title},
% % \emph{J. Abbrev.} {\bf vol} (year) pg.

% % \bibitem{b}
% % Author,
% % \emph{Title},
% % arxiv:1234.5678.

% % \bibitem{c}
% % Author,
% % \emph{Title},
% % Publisher (year).
% \end{thebibliography}
% \bibliographystyle{plainnat}
% \bibliography{reference}
%\bibliographystyle{JHEP}
%\bibliography{reference}
% \bibliographystyle{JHEP}
% \bibliography{reference}
\end{document}